\documentclass{epl}
\newcommand {\dr}{{\mathrm d}\mathbf{r}}

\newcommand {\rr}{\mathbf{r}}
\shorttitle{Solvent mediated long range forces}
\title{Microscopic theory of solvent mediated long range forces: influence of
wetting}
\shortauthor{A.J. Archer \etal}
\author{A.J. Archer\inst{1} \and R. Evans\inst{1} \and R. Roth\inst{2,3}}
\institute{
  \inst{1} H.H. Wills Physics Laboratory, University of Bristol - Bristol BS8
  1TL, UK\\
  \inst{2} Max-Plank Institut f\"ur Metallforschung - Heisenbergstr. 1,
  70569 Stuttgart, Germany\\
  \inst{3} ITAP, Universit{\"a}t Stuttgart - Pfaffenwaldring 57, 70569 
  Stuttgart, Germany}
\begin{document}
\maketitle
\begin{abstract}
We show that a general density functional approach for calculating the force
between two big particles immersed in a solvent of smaller ones can describe
systems that exhibit fluid-fluid phase separation: the theory captures
effects of strong adsorption (wetting) and of critical fluctuations in the
solvent. We illustrate the approach for the Gaussian core model, a simple model
of a polymer mixture in solution and find extremely attractive, long ranged
solvent mediated potentials between the big particles for state points lying
close to the binodal, on the side where the solvent is poor in the species which
is favoured by the big particles.
\end{abstract}
Determining the effective force between two (big) particles immersed in a
solvent constitutes a canonical problem in condensed matter science. From the
statistical mechanics viewpoint one should integrate out the relevant degrees of
freedom of the solvent particles
in order to obtain the the effective pair potential \cite{McMillan}. In
the case of colloidal systems, where the colloidal component is very much larger
than the particles constituting the solvent, the description of the complex,
multi-component system in terms of an effective colloid-colloid potential forms
a cornerstone of the subject. Well-known examples are the DLVO potential for
charge stabilized systems and the hard-sphere potential for sterically
stabilized systems \cite{Likos}. The effective potential between colloids
exhibits new features if particles of intermediate size, such as non-adsorbing
polymer or small colloids, are added to the solvent. When two big colloids ($b$)
are sufficiently close that the intermediate sized particles are
depleted from the region between the big colloids the effective $bb$ potential,
$V_{bb}^{eff}(r)$, can exhibit entropically driven attraction. Within the
Asakura-Oosawa (AO) model \cite{AO}, where the colloids are hard spheres and the
polymer is treated as ideal (inter-penetrating and non-interacting) the
resulting depletion potential is purely attractive and its finite range is equal
to the `diameter' of the polymer coil. For mixtures of hard-sphere
like colloids the effective potential between the bigger colloids exhibits both
attraction and repulsion; short ranged correlations arising from the packing
of the smaller colloids gives rise to an exponentially damped, oscillatory
effective potential whose decay length is simply the bulk correlation length
\cite{Roland}, in keeping with the fact that
$V_{bb}^{eff}(r)=-k_{B} T \ln g_{bb}(r)$, where $T$ is the temperature and
$g_{bb}(r)$ is the $bb$ radial distribution function of the mixture infinitely
dilute in species $b$.

\begin{figure}
\onefigure[width=7cm,height=5.5cm]{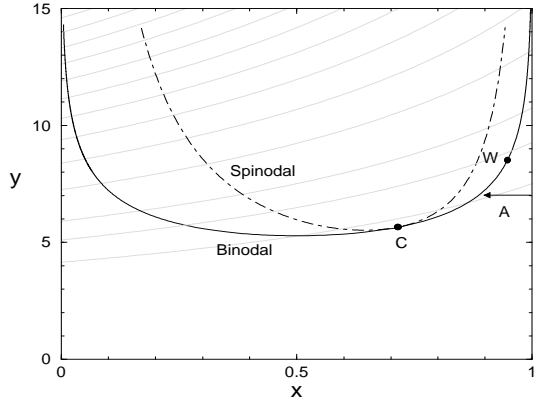}
\caption{Phase diagram of a solvent exhibiting fluid-fluid phase separation.
For gas-liquid phase separation $x$ corresponds to the fluid density and 
$y$ to inverse temperature. In the case of a binary mixture, $x$ is the relative
concentration of one of the species, and $y$ corresponds to the temperature or,
alternatively, to the total bulk 
density, $\rho^0$. For the present binary fluid of GCM particles 
$x=\rho_2^0/\rho^0$, where $\rho_2^0$ is the density of species 2, and 
$y=\rho^0 R_{11}^3$ where $R_{11}$ is the radius of the larger species 1. The
pair potential parameters are: 
$\epsilon_{12}/ \epsilon_{11} =0.944$ and $R_{22}/R_{11}=0.665$ which is 
equivalent to a mixture of two polymers with length ratio 2:1. The gray 
lines are isobars, the 
lowest is at reduced pressure $P \beta R_{11}^3 =100$, the next at 
$P \beta R_{11}^3 =150$, then 200 with the subsequent ones increasing in 
increments of 100. $W$ denotes the wetting point, i.e. the surface phase
transition, below which a thick wetting film of the phase rich in species 1 can
grow on a wall or on a big particle immersed in the binary mixture at a state
near the right hand binodal. Arrow $A$ indicates the path along which the 
density profiles in Fig.~\ref{fig:gaussian_wetting} are calculated. $C$ denotes
the critical point.}
\label{fig:phase_diag}
\end{figure}

In this letter we investigate more generally the influence of correlations in
the solvent on the nature of effective forces between two big particles. We
suppose the solvent has a phase diagram such as that shown in
fig.~\ref{fig:phase_diag}. The main features are: (i) possible phase
separation into coexisting liquid and vapour phases, if the solvent is a pure
system, or into two coexisting fluid phases, if the solvent
is a binary mixture; (ii) a wetting point (denoted $W$ in 
fig.~\ref{fig:phase_diag}) below which macroscopically thick wetting layers of
the coexisting phase can grow on a planar wall upon
approaching coexistence along a path such as that denoted by $A$
\cite{Dietrich,Archer2}; (iii) a critical point $C$ at which the correlations in
the solvent become macroscopically long ranged, i.e.\ the correlation length
diverges.

For states well-removed from the coexistence curve (binodal) correlations in
the solvent are short ranged, and the solvent mediated (SM) forces should also
be short ranged. Qualitatively different features should arise in
$V_{bb}^{eff}(r)$ when long-ranged correlations occur.
There are two obvious mechanisms for such behaviour. First, critical
fluctuations of the solvent should give rise to long-ranged forces between
plates or big particles \cite{Fisher}.
These `critical Casimir forces' are expected to induce flocculation
of colloids suspended in near-critical solvents. The second mechanism, and the
one of primary interest here, is
that associated with the growth of wetting films around plates or sufficiently
large particles. We consider a path such as $A$ in fig.~\ref{fig:phase_diag},
which lies below the wetting point $W$. On approaching the binodal the
coexisting phase (small concentration $x$) will wet completely the interface
between a planar
wall and the bulk fluid phase (large $x$). In the case of large particles, made
of the same material as the wall, thick adsorbed films will develop but,
because of the finite radius of curvature, films will remain of finite thickness
even at coexistence \cite{Dietrich}. One might expect to find long ranged SM
forces arising from the presence of such `wetting' films. Indeed wetting induced
effective potentials between spherical particles immersed in a one-component
fluid have been calculated using an interface displacement (sharp-kink)
description of the solvent density distribution \cite{Bauer}. Although
well-suited to very big particles and thermodynamic states very close to
coexistence, where very thick films can develop, and to systems where
dispersion forces dominate, the approach of ref.~\cite{Bauer} is less
well-suited to situations where the particles are not enormous, so that the
films are thinner, and where dispersion forces do not dominate. Our present
approach, whilst also based on density functional theory (DFT), implements the
general method for including correlation effects developed in the calculation of
depletion potentials \cite{Roland}. Since
it makes no particular assumption about the form of the density distributions
it is applicable for all (fluid) states, including those close to the critical
point where a sharp-kink approximation is inappropriate. We first treat one big
particle as an external potential by fixing its position at the origin and then
calculate the equilibrium density profile(s) of the solvent particles in this
external potential. Thus we input a fully microscopic description of the 
`wetting film' (or of the long range decay of the density profile(s) if we
are near the critical point). The next step is to calculate the SM potential by
inserting a second big particle using the potential distribution theorem
\cite{Roland,Henderson}. To this end we require a functional capable of
describing a mixture of the solvent and big particles that is reliable
in the limit of vanishing density of big particles
\cite{Roland}. There has been progress in this regard. The Rosenfeld
functional \cite{Rosenfeld} for hard sphere mixtures was used successfully to
calculate depletion potentials for a wide range of size ratios \cite{Roland}.
However, hard-sphere mixtures do not exhibit fluid-fluid demixing.
Recently a DFT for the AO model of a colloid-polymer mixture was
derived \cite{Brader}. For certain
size ratios the mixture undergoes phase separation into colloid rich and
colloid poor fluid phases and the DFT predicts entropically driven complete
wetting of a planar hard wall by the colloid rich phase \cite{Brader2}. It is
straightforward to generalize the functional to include an additional component
of big hard spheres at which `wetting' films develop as coexistence is
approached from the colloid poor side of the binodal. Results for effective
potentials between the big hard spheres will be presented
elsewhere \cite{AO_version}.

Another, extremely simple functional \cite{paper1,Archer1} for a mixture of soft
core repulsive Gaussian particles, which models the interaction between the
centers of mass of polymers in solution \cite{Likos}, was also found to be
capable of treating mixtures with large size asymmetries. The important feature
for the present study is that
whilst the one component Gaussian core model (GCM) does not phase separate, a
binary mixture of two different sized GCM particles does separate into
two fluid phases \cite{paper1,Archer1} and displays wetting transitions
\cite{Archer2} for certain, purely repulsive, planar walls. The phase diagram of
fig.~\ref{fig:phase_diag} refers to the particular binary GCM mixture whose
wetting properties were studied in ref.~\cite{Archer2}. Here we consider a
ternary GCM mixture with the density of the biggest (repulsive) Gaussian
particles $\rho_b^0 \rightarrow 0$.

We outline the general method used to calculate the SM potential -- a
more detailed account can be found in ref.~\cite{Roland}. The SM 
potential between two objects $a$ and $b$ is defined by $W_{ab}(\rr_b) \, =
\, \Omega(\rr_b)-\Omega(\rr_b \rightarrow \infty)$, where $\Omega(\rr_b)$ is 
the grand potential of the solvent with object $a$ fixed at $\rr_a=0$ and object
$b$ at $\rr_b$. Using the potential distribution theorem this difference in
grand potential can be written in terms of the one-body direct correlation
function $c_b^{(1)}$ in the limit of the chemical potential of species $b$,
$\mu_b \rightarrow -\infty$, equivalent to the limit of the bulk density,
$\rho_b^0 \rightarrow 0$ \cite{Roland}:
\begin{equation}
\beta W_{ab}(\rr_b) \, = \, c_b^{(1)}(\rr_b \rightarrow\infty; \mu_b
\rightarrow-\infty) -c_b^{(1)}(\rr_b;\mu_b \rightarrow -\infty),
\label{eq:2}
\end{equation}
where $\beta=1/k_BT$. The one-body correlation function $c_b^{(1)}(\rr)$ is
given within DFT \cite{Evans79} by $c_b^{(1)}(\rr) \, = \, -\beta \delta{\cal
F}_{ex}[\{\rho_i \}]/ \delta\rho_b(\rr)$,
where ${\cal F}_{ex}$ is the excess (over ideal gas) Helmholtz free energy
functional of the mixture of the solvent and the big particles, $\rho_i$
denotes the density profile of species $i$, with $\rho_b$ that of the big
particles. The effective potential between
two big particles is then the sum of the bare interaction $v_{bb}(r)$ and the SM
potential: $V_{bb}^{eff}(r)=v_{bb}(r)+W_{bb}(r)$. In order to 
implement eq.~(\ref{eq:2}), an appropriate approximation for ${\cal F}_{ex}$ for
the mixture under consideration must be specified.

We focus on the GCM and refer the 
reader interested in its origins to ref.~\cite{Likos}. The structure of 
the bulk GCM fluid at high densities is very different from that of 
hard core fluids, since the particle cores are soft and hence can overlap. 
For a mixture the pair potential between particles is given by
$\beta v_{ij}(r)=\epsilon_{ij}^*\exp(-r^2/R_{ij}^2)$, where for polymers in
an athermal solvent at room temperature $\epsilon_{ij}^* \simeq 2$ and $R_{ii}$
is roughly the radius of gyration. The radius of cross interaction is usually
taken to be $R_{ij}^2=(R_{ii}^2+R_{jj}^2)/2$ \cite{paper1,Dautenhahn}. The cross
interaction energy parameter $\epsilon_{i \neq j}^*<\epsilon_{ii}^*$ reflects
the way a small polymer coil can overlap with a larger polymer with a smaller
energy penalty than for a polymer of the same size. The particular values we
choose are $\epsilon_{11}^*=\epsilon_{22}^*=2$, $\epsilon_{12}^*=1.89$,
$R_{22}/R_{11}=0.665$ \cite{paper1,Archer1,Archer2} and $\epsilon_{b1}^*=1.0$,
$\epsilon_{b2}^*=0.8$ and $R_{bb}/R_{11}=7.0$, resulting in radii of cross
interaction of $R_{b1}/R_{11}=5.0$ and $R_{b2}/R_{11}=4.97$. At densities where
the soft cores of the GCM particles start to overlap, such that a particle
interacts with a large number of its neighbours, the fluid behaves as a
mean-field fluid \cite{Likos}. The simple random phase approximation (RPA),
which states that the pair direct correlation function is simply
$c_{ij}^{(2)}(r)=-\beta v_{ij}(r)$, accounts well for the bulk correlations at
these high densities \cite{Likos,paper1,Archer1}. The excess free energy
functional which generates the RPA for an arbitrary number of components is
\begin{equation}
{\cal F}_{ex}[\{\rho_i\}]\, =\, \frac{1}{2} \sum_{ij} \int\dr_1\int\dr_2\,\, 
\rho_i(\rr_1)\,\rho_j(\rr_2) v_{ij}(|\rr_1-\rr_2|).
\label{eq:F_ex_GCM}
\end{equation}
since $c_{ij}^{(2)}(\rr_1,\rr_2)= - \beta \delta^2 {\cal F}_{ex}/\delta
\rho_i(\rr_1) \delta \rho_j(\rr_2)$ \cite{Evans79}. Using this (RPA) functional
for the ternary mixture, with $i,j=1,2$ and $b$, in eq.~(\ref{eq:2}) the SM
potential reduces to
\begin{equation}
W_{bb}(\rr_1) \, =\, -\sum_{i=1}^2 \int\dr_2 ( \rho_i(\rr_2) - \rho_i^0 )
v_{bi}(|\rr_1-\rr_2|),
\label{eq:depletion}
\end{equation}
where the sum now runs over the small particle species only since $\rho_b^0
\rightarrow 0$. The profiles $\rho_i(\rr)$, $i=1,2$, are the density profiles of
the small particle species obtained by solving the Euler-Lagrange equations for
the situation where the big particle exerts an external potential on the binary
mixture \cite{Roland}. $\rho_i^0$ are the small particle bulk densities. Near
the binodal (for example on path $A$, fig.~\ref{fig:phase_diag}), the 
density profiles of the smaller particle species will reflect the formation of a
thick adsorbed film around a single big particle, and so within the RPA
treatment we have a prescription
for including at least some of the effects of wetting on the SM potential.

\begin{figure}
\twofigures[width=7cm,height=5.5cm]{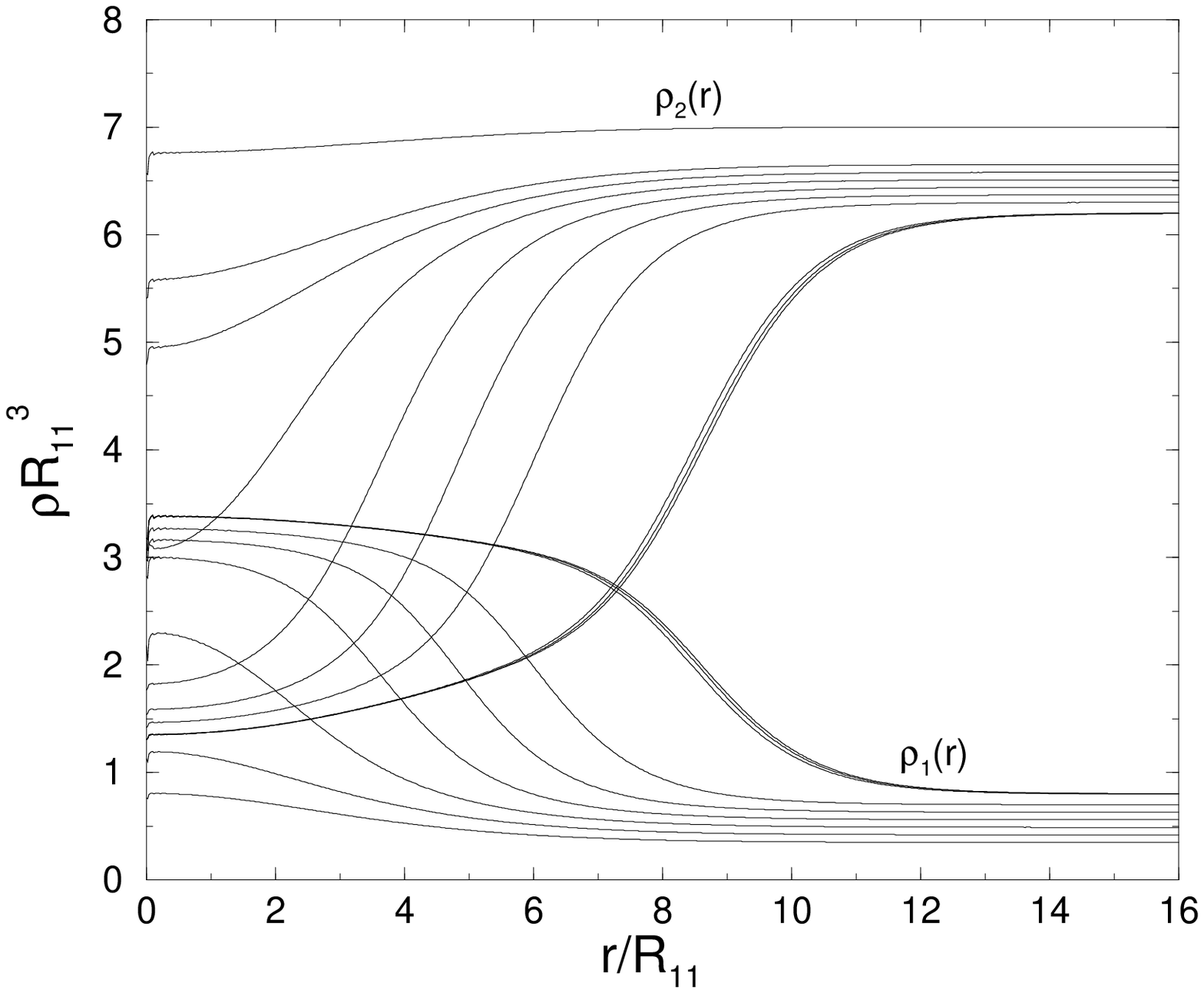}{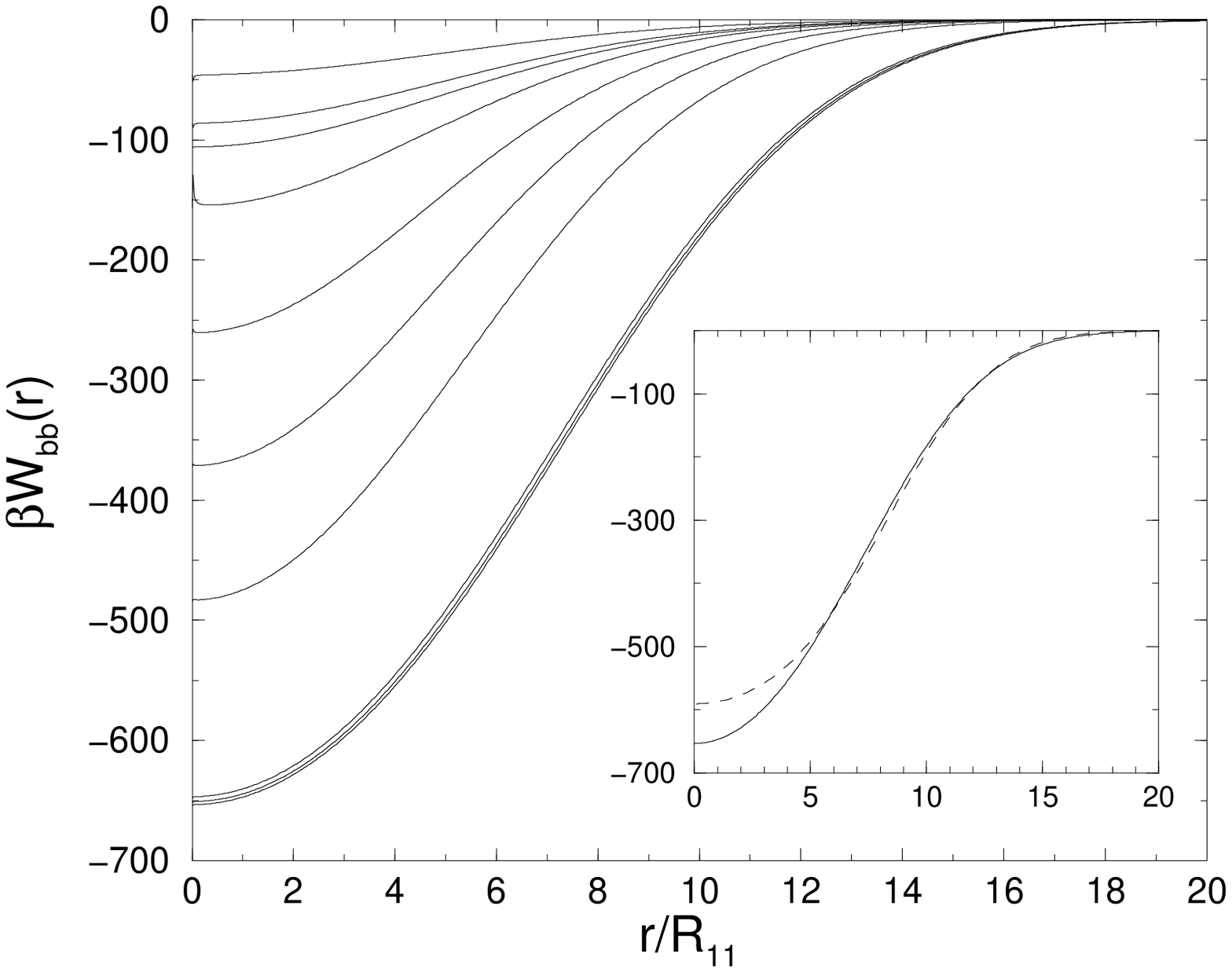}
\caption{The density profiles of a binary solvent of small GCM particles, of
radii $R_{11}$ and $R_{22}$ with $R_{22}/R_{11}=0.665$, around a large GCM 
test particle with $R_{bb}/R_{11}=7.0$, calculated at constant total bulk 
density $\rho^0 R_{11}^3=7.0$ i.e.\ along path $A$ in fig.~\ref{fig:phase_diag},
at concentration
$x=\rho_2^0/\rho^0=1.0$, 0.95, 0.94, 0.93, 0.92, 0.91, 0.9, 0.886, 0.8857 and
0.8855 ($x_{coex.}=0.8854$). The top set of profiles are those of the smaller 
of the two species, species 2. The very top one is for $x=1.0$ and the one below
is for $0.95$ etc. The set of profiles at the bottom correspond to species 1;
the very lowest profile is at $x=0.95$ (for $x=1.0$ the profile is
$\rho_1(r)=0$), the one above is for 0.94 etc. Note the pronounced change in
shape of the density profiles from roughly Gaussian away from coexistence, to a
non-Gaussian shape, with a flat portion near the origin and the free
interface-like `knee', close to coexistence. The latter shape signals a thick
adsorbed film of the coexisting phase, rich in species 1, growing around the big
GCM particle.}
\label{fig:gaussian_wetting}
\caption{The SM potential between two big GCM particles with radius
$R_{bb}/R_{11}=7.0$ as calculated from the density profiles in
Fig.~\ref{fig:gaussian_wetting}. The
potentials correspond to concentration 
$x=1.0$, 0.95, 0.94, 0.93, 0.92, 0.91, 0.9, 0.886, 0.8857 and
0.8855, going from top to bottom ($x_{coex.}=0.8854$). The SM potential
increases in depth and range as the binodal is approached. The
energy scale close to coexistence is set by the difference in the grand potential
between inserting a single big particle in the bulk and inserting it into the
bulk coexisting phase which constitutes the adsorbed film and for the GCM (see
text) this results in particularly deep
potentials. In the inset we re-plot the SM potential calculated 
at $x=0.8855$ (solid line) along with our analytic approximation,
eq.~(\ref{eq:sharp_kink_dep}) (dashed line), with $l/R_{11}=9.5$.}
\label{fig:dep_2comp}
\end{figure}

We study the same binary mixture of GCM particles, equivalent to a binary
mixture of polymers of length ratio 2:1, as in refs.~\cite{Archer1,Archer2}.
Into this phase separating binary mixture we now add the third, much bigger
species of GCM particle with $R_{bb}/R_{11}=7.0$. In
fig.~\ref{fig:gaussian_wetting} we show the density profiles of both components
of the binary solvent 
calculated along path $A$ in fig.~\ref{fig:phase_diag}, where the total bulk 
density is constant, $\rho^0 R_{11}^3=7.0$. As the the binodal is approached 
the shape of the profiles changes dramatically as the thick adsorbed film
develops. For concentrations $x$ close to the coexistence value, $x_{coex}$, the
profiles have a flat central portion where the densities take values roughly
similar to those in the coexisting bulk phase rich in species 1. We denote these
$\rho_i^{coex}$, $i=1,2$. In this regime the adsorbed film extends beyond the
radius $R_{bb}$ of the big particle fixed at the origin but because of
curvature it remains of finite extent at $x=x_{coex}$. In the opposite regime,
near $x=1$, the profiles are Gaussian-like. Indeed in the limit of a pure
solvent of GCM particles of species 2 the profile can be approximated extremely
accurately by the simple ansatz $\rho_2(r)=\rho_2^0-\rho^* \exp(-r^2/R_{b2}^2)$
with $\rho^* = \rho_2^0 \epsilon_{b2}^*/(1+\rho_2^0 \pi^{3/2} \epsilon_{22}^*
R_{22}^3)$, provided $R_{bb}/R_{22} \gg 1$ \cite{Archer3}. When this ansatz is
inserted into eq.~(\ref{eq:depletion}) we find that
\begin{equation}
\beta W_{bb}^{pure}(r) \,=\, -(\pi/2)^{3/2} \epsilon_{b2}^* \rho^* R_{b2}^3
\exp(-r^2/2R_{b2}^2),
\label{eq:ansatz1}
\end{equation}
which lies on top of the full numerical result (top curve in
fig.~\ref{fig:dep_2comp}). Thus even when there is no adsorbed film the SM
potential is strongly attractive ($W_{bb}(r=0) \simeq -45k_BT$ for
$\rho_2^0 R_{11}^3=7.0$). The large amplitude arises from the factor
$\rho^*R_{b2}^3$ in eq.~(\ref{eq:ansatz1}), which is roughly the number of small
particles expelled from the volume of the big one. This factor is large in these
high density states. As we move along path $A$
(fig.~\ref{fig:phase_diag}), adding more of species 1 to the host fluid, we find
that as the binodal is approached the SM potential, obtained by inserting the
profiles of fig.~\ref{fig:gaussian_wetting} into eq.~(\ref{eq:depletion}),
becomes longer ranged and deeper (see fig.~\ref{fig:dep_2comp}). For
example, when $x=\rho_2^0/\rho^0=0.887$ then
$W_{bb}(r=2R_{bb}) \simeq -30k_BT$ and $W_{bb}(r=0) \simeq -650k_BT$. Note that
the underlying (bare) repulsive big-big potential $v_{bb}(r)$ will be negligible
in comparison with such strongly attractive SM potentials. It is
clear that enormous attractive interactions are generated in this mixture where
all the bare interparticle potentials are purely repulsive.

Why does the presence of thick adsorbed films induce such deep
SM potentials? In order to obtain some insight we can approximate the
density profiles of the solvent using the so-called sharp-kink (sk)
approximation \cite{Dietrich}, i.e. $\rho_i(r)=\rho_i^{coex}$ for $0<r \leq l$
and $\rho_i(r)=\rho_i^0$ for $r > l$, where $l$ is the  thickness of the
adsorbed film. On substituting this approximation for $\rho_i(r)$ into
eq.~(\ref{eq:depletion}) we find that
\begin{equation}
\beta W_{bb}^{sk}(r) \, =\,
\frac{\pi}{2}\sum_{i=1}^2\Delta \rho_i\epsilon_{bi}^*R_{bi}^3
\left\{ \sqrt\pi \left[\mathrm{erf} \left( r_- \right) 
+\mathrm{erf} \left( r_+ \right) \right]
-\frac{R_{bi}}{r}(e^{-r_-^2}
-e^{-r_+^2})\right\},
\label{eq:sharp_kink_dep}
\end{equation}
where $\Delta \rho_i=(\rho_i^{coex}-\rho_i^0)$ is the difference in coexisting
density for species $i$, $\mathrm{erf}(x)=(2/\sqrt\pi)\int_0^x dt \exp(-t^2)$ is
the error function and $r_\pm = (l \pm r)/R_{bi}$. The energy scale at
coexistence is set by $\beta \Delta \Omega \equiv
c_b^{(1)}(\infty;\{\rho_i^{coex}\}) - c_b^{(1)}(\infty; \{\rho_i^0\}) = -\sum_i
\Delta \rho_i \pi^{3/2} \epsilon_{bi}^* R_{bi}^3 >0$
\cite{Archer3}, the difference in the grand potential between inserting a single
large GCM particle in the bulk fluid near the phase boundary (bulk densities
$\rho_i^0$) and inserting the large particle in the bulk coexisting phase (the
phase adsorbing on the big particle, densities $\rho_i^{coex}$). For a thick 
enough adsorbed film around the big particle, there will be a region inside the
film where we can insert the second big particle sufficiently far 
away from both the fluid-fluid interface and from the central big particle at
the origin,
that the grand potential for inserting the big particle is approximately that of
inserting it into the bulk of the coexisting phase. For the point on the binodal
at $\rho^0 R_{11}^3=7.0$ (the intersection with path $A$
fig.~\ref{fig:phase_diag}) we find $\beta \Delta \Omega =627$, the large value
arising mainly from the high values of $(R_{bi}/R_{11})^3$, which should be a
general feature of the GCM. The comparison between the sharp
kink approximation and the full calculation is shown in the
inset to fig.~\ref{fig:dep_2comp}, taking a film thickness $l/R_{11}=9.5$. We
find that eq.~(\ref{eq:sharp_kink_dep}) captures the gross features of the SM
potential calculated numerically for $x$ very close to $x_{coex}$.
In other types of fluids the quantity $\beta \Delta \Omega$ should also set the
energy scale at coexistence but its form will depend on the particular fluid.
For example in the AO model the depth of the depletion potential scales roughly
as $R_{cc}/R_{pp}$, where $c$ refers to the hard-sphere colloid and $p$ to ideal
polymer. We expect crossover to $(R_{bb}/R_{pp})^2$ scaling when two very big
hard-spheres are immersed in the colloid poor phase and are `wet' by the colloid
rich phase at coexistence \cite{AO_version}.

We have also calculated the SM potential for states in the vicinity of the
critical point $C$ in fig.~\ref{fig:phase_diag}. When the bulk correlation
length of the binary GCM is several times $R_{11}$, $W_{bb}(r)$ decays slowly
with separation $r$, i.e.\ the SM force becomes very long ranged. Results will
be reported elsewhere \cite{Archer3}. Note that our general procedure is
formally equivalent to calculating the effective $bb$ potential from the
logarithm of $g_{bb}(r)$, in the limit $\rho_b^0 \rightarrow 0$, via the test
particle route \cite{Roland}. Since the procedure can also be used to
calculate the SM potential between a test particle and a planar wall (or another
fixed object) \cite{Roland} we can investigate the influence of wetting films on
effective wall-particle forces for arbitrary sizes of test particle.

Some impetus for theories \cite{Bauer} of wetting-induced interaction potentials
comes from experimental observations of striking flocculation phenomena for
colloidal particles immersed in a binary liquid mixture close to the binodal
\cite{experiments}. Flocculation appears to take place on the side of the phase
diagram where the mixture is poor in the species which is favoured by the
colloid. This is where we predict strongly attractive forces. Although the
precise interpretation of the experiments remains somewhat controversial --
critical Casimir forces and other mechanisms specific to the experimental
systems might also play a role -- it is clear that wetting can have a profound
effect on effective colloid-colloid interactions.

The approach we have outlined here is a step towards a general microscopic
theory of solvent mediated interactions which can incorporate the effects of
both wetting and bulk criticality, albeit at a mean-field level, for a wide
range of particle sizes. As such it complements the mesoscopic approach of
ref.~\cite{Bauer} and some other approaches used in studies
of near-critical solvents \cite{Schlesener}. We find that for a particular
model, namely a mixture of repulsive Gaussian core particles, the SM
interactions between the big particles can be long ranged and very strongly
attractive near the binodal. We expect to find similar long-ranged attractive
forces induced by
wetting in all solvents exhibiting fluid-fluid phase separation.

\acknowledgments
We benefitted from helpful discussions with J.M. Brader, M. Schmidt, and H.
L{\"o}wen. AJA was supported by an EPSRC studentship.


\begin{thebibliography}{0}

\bibitem{McMillan}
  \Name{W.G McMillan \and J.E. Mayer}
  \REVIEW{J. Chem. Phys.}{13}{1945}{276}.

\bibitem{Likos}
For a recent review of effective interactions in colloidal and other soft matter
systems see
  \Name{C.N. Likos}
  \REVIEW{Phys. Reports}{348}{2001}{267}.
For charged colloids see
  \Name{J-.P. Hansen \and H. L\"owen}
  \REVIEW{Ann. Rev. Phys. Chem.}{51}{2000}{209}.

\bibitem{AO}
  \Name{S. Asakura \and F. Oosawa}
  \REVIEW{J. Chem. Phys.}{22}{1954}{1255}.

\bibitem{Roland}
  \Name{R. Roth, R. Evans, \and S. Dietrich}
  \REVIEW{Phys. Rev. E}{62}{2000}{5360}
and references therein.

\bibitem{Dietrich}
see e.g.\
  \Name{S. Dietrich}
  \Book{Phase Transitions and Critical Phenomena}
  \Editor{C. Domb \and J.L. Lebowitz}
  \Vol{12}
  \Publ{London: Academic}
  \Year{1988}
  \Page{1}.
  
\bibitem{Archer2}
  \Name{A.J. Archer \and R. Evans}
  \REVIEW{J. Phys.: Condens. Matter}{14}{2002}{1131}.

\bibitem{Fisher}
  \Name{M.E. Fisher \and P.G. de Gennes}
  \REVIEW{C.R. Acad. Sc. Paris B}{287}{1978}{207}.

\bibitem{Bauer}
  \Name{C. Bauer, T. Bieker, \and S. Dietrich}
  \REVIEW{Phys. Rev. E}{62}{2000}{5324}.

\bibitem{Henderson}
  \Name{J.R. Henderson}
  \REVIEW{Mol. Phys.}{50}{1983}{741}.

\bibitem{Rosenfeld}
  \Name{Y. Rosenfeld}
  \REVIEW{Phys. Rev. Lett.}{63}{1989}{980}.

\bibitem{Brader}
  \Name{M. Schmidt, H. L\"owen, J.M. Brader \and R. Evans}
  \REVIEW{Phys. Rev. Lett.}{85}{2000}{1934}.

\bibitem{Brader2}
  \Name{J.M. Brader, R. Evans, M. Schmidt \and H. L\"owen}
  \REVIEW{J. Phys. Cond. Matt.}{14}{2002}{L1}.

\bibitem{AO_version}
  \Name{R. Roth, \and R. Evans}
  in preparation.

\bibitem{paper1}
  \Name{A.A. Louis, P.G. Bolhuis, \and J-P. Hansen}
  \REVIEW{Phys. Rev. E}{62}{2000}{7961}.

\bibitem{Archer1}
  \Name{A.J. Archer \and R. Evans}
  \REVIEW{Phys. Rev. E}{64}{2001}{041501}.

\bibitem{Evans79}
  \Name{R. Evans}
  \REVIEW{Adv. Phys}{28}{1979}{143}.

\bibitem{Dautenhahn} 
  \Name{J. Dautenhahn \and C.K. Hall}
  \REVIEW{Macromolecules}{27}{1994}{5399}.

\bibitem{Archer3}
  \Name{A.J. Archer \and R. Evans}
  in preparation.

\bibitem{experiments}
Many of these experiments were carried out by D. Beysens and co-workers. For a
comprehensive list of references and a summary of attempts to explain the data
see 
  \Name{T. Bieker, \and S. Dietrich}
  \REVIEW{Physica A}{252}{1998}{85}
and ref.~\cite{Bauer}.

\bibitem{Schlesener} 
  \Name{F. Schlesener, A. Hanke \and S. Dietrich}
{\em cond-mat/0202532}, and references therein.

\end{thebibliography}
\end{document}